\documentclass[aps,prc,twocolumn,superscriptaddress,showpacs,amsmath,floatfix,
nofootinbib]{revtex4}

\usepackage{graphicx}
\usepackage{bm}

\begin{document}


\title{Particle-number conservation in static-path approximation for thermal 
superfluid systems}

\author{K. Kaneko}
\email{kaneko@ip.kyusan-u.ac.jp}
\affiliation{Department of Physics, Kyushu Sangyo University, Fukuoka 813-8503,
Japan}
\author{A. Schiller}
\affiliation{National Superconducting Cyclotron Laboratory, Michigan State 
University, East Lansing, MI 48824}

\begin{abstract}
By applying particle-number projection to the static-path approximation (SPA), 
the heat capacity and the breakdown of pairing correlations are investigated in
the thermally excited, superfluid systems $^{172}$Yb, $^{94}$Mo, and $^{56}$Fe.
For the heavy nucleus $^{172}$Yb, the heat capacities in both the SPA and the 
number-projected SPA (NPSPA) exhibit an $\cal{S}$ shape; the difference between
the SPA and NPSPA heat-capacity curves is not very large and the 
particle-number projection thereby enhances the $\cal{S}$ shape already seen in
the SPA\@. The temperature at which the $\cal{S}$-shape of heat capacity curve 
occurs parallels the temperature of the breakdown of pairing correlations as 
indicated by the effective pairing gap. However, for the comparatively lighter 
nuclei $^{94}$Mo and $^{56}$Fe, the SPA does not produce an $\cal{S}$-shaped 
heat capacity on its own; only after particle-number projection the $\cal{S}$ 
shape appears in the heat-capacity curve. For $^{94}$Mo, we compare the NPSPA 
result with thermal odd-even mass differences, which are regarded as a direct 
measure of the pairing gap. 
\end{abstract}


\maketitle


\section{Introduction}

Pairing correlations are of special importance for ma\-ny-fermion systems such 
as electrons in superconducting metals, nucleons in the nucleus, and quarks in 
color superconductivity. The Bardeen-Cooper-Schrieffer (BCS) theory \cite{BCS} 
of superconductivity has successfully described the sharp superfluid-to-normal 
phase transition connected to the breakdown of pairing correlations for an 
infinite Fermi system. Such a second-order phase transition is characterized by
a discontinuity of the second derivative of the partition function with respect
to some order parameter, e.g., in the present case, a jump of the heat capacity
at a critical temperature. For finite Fermi systems, however, recent 
theoretical approaches \cite{Rossignoli,Rombouts,Liu,Alhassid} demonstrate that
thermal and quantal fluctuations are important; they wash out the discontinuity
of the heat capacity which is obtained in the BCS approximation, and instead 
produce a continuous $\cal{S}$-shape around the critical temperature. Such an 
$\cal{S}$-shape has been found experimentally by the Oslo group 
\cite{Schiller,Melby}, and interpreted as a signature of the pairing phase 
transition. Another fingerprint of this transition is the local decrease of 
thermal odd-even mass differences \cite{Kaneko,Kaneko1,Kaneko2} extracted from 
the observed level densities of a triplet of isotopes with neutron number 
$N-1$, $N$, and $N+1$, which yields a temperature corresponding to the one 
obtained from the $\cal{S}$ shape of the heat capacity curve. 

In finite Fermi systems such as nuclei, the BCS theory alone fails to provide a
good approximation of thermal properties. This comes because particle number is
not a good quantum number in the BCS description of the superfluid phase. 
Indeed, a BCS treatment with rigorous number projection \cite{Essebag} such as 
the variation after projection (VAP) method \cite{Essebag,Nakada} smooths out 
the sharp phase transition and hence gives a more realistic picture, while the 
discontinuity in the heat capacity remains in, e.g., the projection after 
variation (PAV) method \cite{Essebag,Rossignoli2,Tanabe,Tanabe1}\@. Exact 
number projection is also essential for an accurate description of odd-even 
effects seen in the heat-capacity curves of small superfluid systems. It is 
therefore desirable to investigate how the number projection affects thermal 
properties such as the $\cal{S}$ shape in heat capacity of nuclei 
\cite{Nakada,Esashika}\@. In particular, it has been recently demonstrated 
\cite{Esashika}, that particle-number projection within the BCS theory by the 
PAV scheme can produce an $\cal{S}$-shaped heat capacity even in the absence of
a pairing-phase transition, i.e., when assuming a constant pairing gap at all 
temperatures due to the effect of particle-number conservation on 
quasi-particle excitations. In general, this method produces both an 
${\cal{S}}$-shaped heat capacity due to the effect of particle-number 
projection and a discontinuity in the heat capacity related to the pairing 
phase transition. 

Inclusion of fluctuations and correlations beyond the BCS theory as induced by,
e.g., exact particle number projection, can be done starting with the 
path-integral representation of the partition function. A direct approach to 
evaluate the path integral is the shell-model Monte-Carlo method 
\cite{Rombouts,Liu,Alhassid}\@. However, the computational effort is quite 
large, and it cannot be applied to the large shell-model spaces typical for 
heavy nuclei. The static-path approximation (SPA) 
\cite{Alhassid1,Bertsch,Rossignoli1,Rossignoli2} is therefore a useful 
treatment to evaluate approximately the partition function in finite systems 
with separable interactions. In recent theoretical approaches, also 
small-amplitude fluctuations around the static path have been taken into 
account. These fluctuations give corrections to the partition function similar 
to the standard random-phase approximation around the mean field. Thus, the 
static-path plus random-phase approximation (SPA+RPA) method for interactions 
and temperature regions applied in \cite{Puddu,Lauritzen,Rossignoli,Attias} 
gives excellent agreement with exact results \cite{Rossignoli}\@. In this 
paper, we will perform the exact number projected SPA (NPSPA), since the SPA 
results for the $\cal{S}$-shaped heat capacity qualitatively agree with the 
SPA+RPA ones \cite{Kaneko3}\@. For this, we employ thermo field dynamics (TFD)
\cite{Takahashi,Umezawa}, which is a powerful tool for describing many-body 
systems at finite temperature. 

\section{Method}

In this work, we consider a monopole pairing Hamiltonian 
\begin{equation}
\hat{H}=\sum_{k,\tau}\varepsilon_{k,\tau}\left(\hat{c}_{k,\tau}^\dagger
\hat{c}_{k,\tau}+\hat{c}_{\bar{k},\tau}^\dagger\hat{c}_{\bar{k},\tau}\right)-
\sum_{\tau}G_\tau\hat{P}_\tau^\dagger\hat{P}_\tau,
\label{eq:1}
\end{equation}
with the time reversed states $\bar{k}$, and the pairing force strength 
$G_\tau$, where $\tau=\pi,\ \nu$ stands for protons and neutrons respectively. 
Here, $\varepsilon_{k,\tau}$ are the single-particle energies and 
$\hat{P}_\tau$ is the pairing operator 
$\hat{P}_\tau=\sum_k\hat{c}_{\bar{k},\tau}\hat{c}_{k,\tau}$\@. 
\footnote{Hereafter, the index $\tau$ is dropped for convenience.}

By means of the number-projected SPA \cite{Rossignoli,Rossignoli2} based on the
Hubbard-Stratonovich transformation \cite{Hubbard}, the canonical partition 
function is given by 
\begin{eqnarray}
Z_N&=&{\mathrm{Tr}}\,\left[\hat{P}_Ne^{-\beta\hat{H}^\prime}
\right]_{\mathrm{SPA}}
\nonumber\\
&=&\frac{2}{GT}\int_0^\infty{\mathrm{d}}\Delta\,\Delta\,e^{-\Delta^2/GT}
Z_N(\Delta),
\label{eq:2}
\end{eqnarray}
with 
\begin{eqnarray}
Z_N(\Delta)&=&{\mathrm{Tr}}\,\left[\hat{P}_Ne^{-\beta\hat{h}(\Delta)}\right],
\label{eq:3}\\
\hat{h}(\Delta)&=&\sum_k\varepsilon_k^\prime\left(\hat{c}_k^\dagger\hat{c}_k+
\hat{c}_{\bar{k}}^\dagger\hat{c}_{\bar{k}}\right)\nonumber\\
&&-\Delta\left(\hat{P}^\dagger+\hat{P}\right)+\frac{G\,\Omega}{2},
\label{eq:4}
\end{eqnarray}
where $\hat{H}^\prime=\hat{H}-\mu\hat{N}$, 
$\hat{N}=\sum_k\hat{c}_k^\dagger\hat{c}_k$ is the particle-number operator, and
$\mu$ is the chemical potential. Furthermore, $\hat{P}_N$ is the exact number 
projection defined by 
\begin{equation}
\hat{P}_N=\frac{1}{2\pi}\int_0^{2\pi}{\mathrm{d}}\varphi\,
e^{-i\varphi(\hat{N}-N)},
\label{eq:5}
\end{equation}
and we define for later $\lambda_k=\sqrt{\varepsilon_k^{\prime 2}+\Delta^2}$ 
with $\varepsilon_k^\prime=\varepsilon_k-\mu- G/2$\@. Then, the thermal energy 
can be calculated from $E=-\partial\ln Z_N/\partial\beta$\@. It is now 
convenient to introduce quasiparticles by diagonalizing the Hamiltonian of Eq.\
(\ref{eq:4}) 
\begin{equation}
\left(\begin{array}{c}\hat{a}_k\\\hat{a}_k^\dagger\end{array}\right)
={\cal{W}}^\dagger
\left(\begin{array}{c}\hat{c}_k\\\hat{c}_k^\dagger\end{array}\right)
=\left(\begin{array}{cc}u_k^\ast&v_k^\ast\\v_k&u_k\end{array}\right)
\left(\begin{array}{c}\hat{c}_k\\\hat{c}_k^\dagger\end{array}\right),
\label{eq:6}
\end{equation}
where the matrix satisfies unitarity ${\cal{W}}^\dagger{\cal{W}}=1$\@. The 
matrices $u$ and $v$ in Eq.\ (\ref{eq:6}) diagonalize the pairing term of Eq.\ 
(\ref{eq:4})\@. They are diagonal and determined by solving the 
Hartree-Fock-Bogoliubov (HFB) equations 
\begin{equation}
\left(\begin{array}{cc}
\varepsilon_k^\prime&\Delta\\\Delta&-\varepsilon_k^\prime\end{array}\right)
\left(\begin{array}{c}u_k\\v_k\end{array}\right)
=\lambda_k\left(\begin{array}{c}u_k\\v_k\end{array}\right),
\label{eq:7}
\end{equation}
where $\lambda_k$ is the quasiparticle energy. 

To evaluate the partition function $Z_N(\Delta)$ in Eq.\ (\ref{eq:3}), we 
employ the treatment of number projection \cite{Tanabe} using the TFD formalism
\cite{Takahashi,Umezawa}\@. The TFD is known to be a powerful tool for a 
perturbative treatment within thermal mean-field theory \cite{Tanabe2}\@. The 
thermal expectation value of an observable is thereby expressed in terms of a 
vacuum expectation value in an enlarged space. The advantage of the TFD is that
the thermal average of an arbitrary operator can be easily handled in a similar
manner as the expectation value in the zero-temperature formalism. In TFD, the 
Hilbert space spanned by the quasiparticle operators is doubled by including a 
fictitious (tilde) operator 
$\{\hat{\tilde{a}}_k,\hat{\tilde{a}}_k^\dagger\}$\@. Then, the quasiparticle 
vacuum is doubled by the corresponding vacuum $|0\rangle$ defined as 
$\hat{a}_k|0\rangle=\hat{\tilde{a}}_k|0\rangle=0$\@. Next, we introduce the 
temperature-dependent operators and vacuum by the unitary transformation 
\begin{eqnarray}
\hat{a}_k(\beta)&=&e^{-i\hat{G}}\hat{a}_ke^{i\hat{G}},
\label{eq:8}\\
\hat{\tilde{a}}_k(\beta)&=&e^{-i\hat{G}}\hat{\tilde{a}}_ke^{i\hat{G}},
\label{eq:9}\\
|0(\beta)\rangle&=&e^{-i\hat{G}}|0\rangle,
\label{eq:10}
\end{eqnarray}
where the generator $\hat{G}$ is given by 
\begin{equation}
\hat{G}=i\sum_k\theta_k\left(\hat{a}_k^\dagger\hat{\tilde{a}}_k^\dagger-
\hat{\tilde{a}}_k\hat{a}_k\right).
\label{eq:11}
\end{equation}
Here, $\theta_k$ is the angle of the transformation. The Fock space is spanned 
by the set of operators 
$\{\hat{a}_k^\dagger(\beta),\hat{\tilde{a}}_k^\dagger(\beta)\}$ and the vacuum 
$|0(\beta)\rangle$\@. The transformations in Eqs.\ (\ref{eq:8}) and 
(\ref{eq:9}) can also be rewritten as 
\begin{equation}
\left(\begin{array}{c}\hat{a}_k\\\hat{\tilde{a}}_k^\dagger\end{array}\right)
=\left(\begin{array}{cc}\cosh\theta_k&\sinh\theta_k\\
\sinh\theta_k&\cosh\theta_k\end{array}\right)
\left(\begin{array}{c}
\hat{a}_k(\beta)\\\hat{\tilde{a}}_k^\dagger(\beta)\end{array}\right).
\label{eq:12}
\end{equation}
With this, the thermal average of an arbitrary operator $\hat{A}$ is expressed 
as the expectation value with respect to the temperature-dependent vacuum 
$|0(\beta)\rangle$ by 
\begin{eqnarray}
\langle\hat{A}\rangle&=& {\mathrm{Tr}}\,\left(\hat{A}e^{-\beta\hat{H}}\right)
\left/\,{\mathrm{Tr}}\,\left(e^{-\beta\hat{H}}\right)\right.\nonumber\\
&=&\langle0(\beta)|\hat{A}|0(\beta)\rangle,
\label{eq:13}
\end{eqnarray}
where $\hat{H}$ is the full Hamiltonian of the system. To determine the angle 
parameter $\theta_k$, we evaluate the thermal average of the number operator 
$\hat{a}_k^\dagger\hat{a}_k$ with respect to $|0(\beta)\rangle$ as 
\begin{equation}
\langle0(\beta)|\hat{a}_k^\dagger\hat{a}_k|0(\beta)\rangle=\sinh^{2}\theta_k.
\label{eq:14}
\end{equation}
Since the left-hand of Eq.\ (\ref{eq:14}) should give the Fermi distribution 
function, the angle is chosen as 
\begin{equation}
f_k=\sinh^2\theta_k=\frac{1}{e^{\beta\lambda_k}+1}.
\label{eq:15}
\end{equation}

\begin{figure}[t!]
\includegraphics[totalheight=8.4cm]{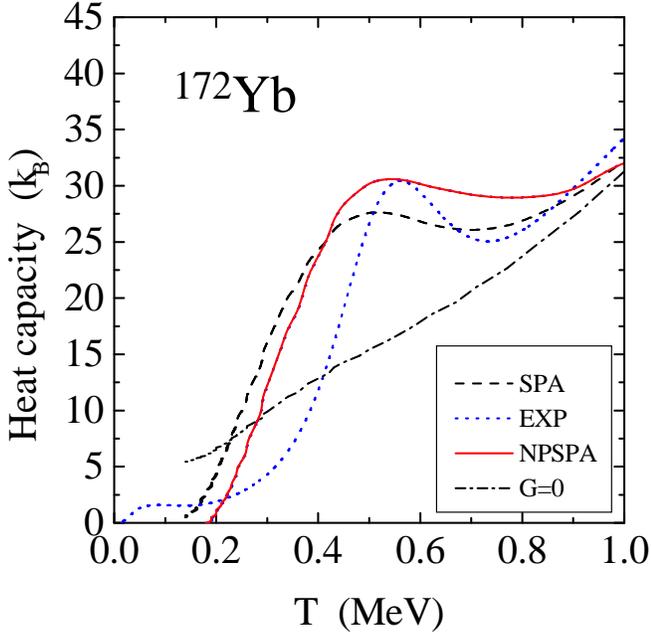}
\caption{(Color online) Heat capacity as a function of temperature for 
$^{172}$Yb. The dashed and solid curves denote results from the SPA and NPSPA, 
respectively, the dash-dotted line gives the result for the 
independent-particle model, i.e., for $G=0$\@. Experimental data (dotted line) 
are taken from \protect\cite{Schiller}\@, Good qualitative agreement between 
the experiment and the calculations is achieved. The temperature around the 
local maximum of the $C_V$ curve coincides well between the experiment and the 
NPSPA\@.}
\label{fig1}
\end{figure}

Using the definition of the thermal average from Eq.\ (\ref{eq:13}), the 
partition function $Z_{N}(\Delta)$ in Eq.\ (\ref{eq:3}) can be rewritten as 
\begin{eqnarray}
Z_N(\Delta)&=&{\mathrm{Tr}}\,\left[\hat{P}_Ne^{-\beta\hat{h}(\Delta)}\right]
\nonumber\\
&=&{\mathrm{Tr}}\,\left(e^{-\beta\hat{h}(\Delta)}\right)\langle0(\beta)|
\hat{P}_N|0(\beta)\rangle,
\label{eq:16}
\end{eqnarray}
with 
\begin{equation}
{\mathrm{Tr}}\,\left(e^{-\beta\hat{h}(\Delta)}\right)=\frac{1}{2}\prod_k
e^{-(\varepsilon_k-\mu-\lambda_k)/T}\left(1+e^{-\lambda_k/T}\right)^2.
\label{eq:17}
\end{equation}
The expectation value $\langle0(\beta)|\hat{P}_N|0(\beta)\rangle$ can be 
evaluated using the TFD technique \cite{Tanabe} in a way similar as in the 
zero-temperature formalism \cite{Hara}\@. For instance, by applying the general
formalism of projection \cite{Tanabe}, we can obtain the expectation value of 
the operator $e^{-i\hat{S}}$ by 
\begin{eqnarray}
\lefteqn{\langle0(\beta)|e^{-i\hat{S}}|0(\beta)\rangle=}\nonumber\\
&&(\det U)^{\frac{1}{2}}\exp\left[-i\left(S^{(0)}+\frac{1}{2}{\mathrm{Tr}}\,
S^{(1)}\right)\right],
\label{eq:18}
\end{eqnarray}
where $\hat{S}$ is defined as the one-body operator $\hat{S}=\varphi\hat{N}$ in
the quasiparticle representation 
\begin{equation}
\hat{S}=S^{(0)}+\sum_{kk^\prime}S_{kk^\prime}^{(1)}\hat{a}_k^\dagger
\hat{a}_{k^\prime}+\frac{1}{2}\sum_{kk^\prime}\left[S_{kk^\prime}^{(2)}
\hat{a}_k^\dagger\hat{a}_{k^\prime}^\dagger+{\mathrm{h.c.}}\right],
\label{eq:19}
\end{equation}
with 
\begin{eqnarray}
S^{(0)}&=&\langle0(\beta)|\hat{S}|0(\beta)\rangle
\label{eq:20}\\
S_{kk^\prime}^{(1)}&=&\langle0(\beta)|[\hat{a}_k,\hat{S}]
\hat{a}_{k^\prime}^\dagger|0(\beta)\rangle
\label{eq:21}\\
S_{kk^\prime}^{(2)}&=&\langle 0(\beta)|\hat{a}_k[\hat{a}_{k^\prime},\hat{S}]
|0(\beta)\rangle,
\label{eq:22}
\end{eqnarray}
and where $U$ is given by the transformation 
\begin{equation}
\exp(-i\hat{S})\hat{a}_k\exp(i\hat{S})=\sum_{kk^\prime}\left(U_{kk^\prime}
\hat{a}_{k^\prime}+V_{kk^\prime}\hat{a}_{k^\prime}^\dagger\right).
\label{eq:23}
\end{equation}

\begin{figure}[t!]
\includegraphics[totalheight=8.5cm]{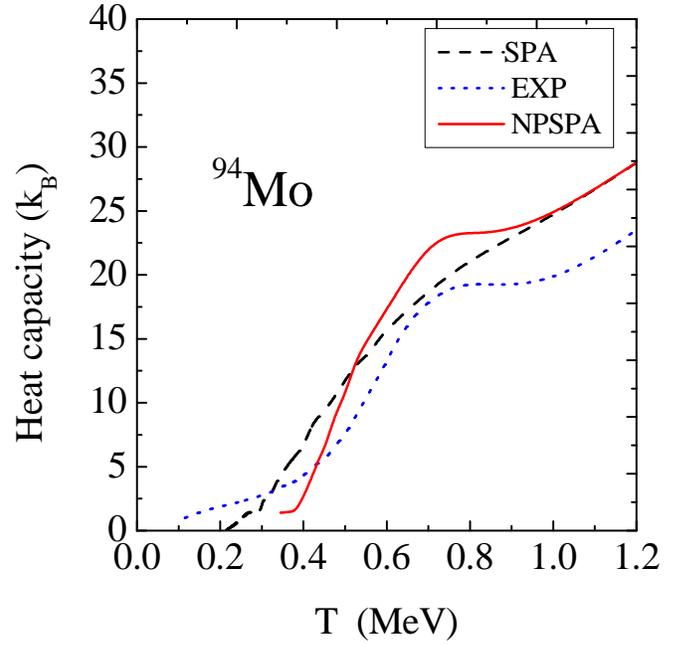}
\caption{(Color online) Heat capacity as a function of temperature for 
$^{94}$Mo. Solid and dashed curves as in Fig.\ \protect\ref{fig1}\@. 
Experimental data (dotted line) are taken from \protect\cite{Chankova}\@. There
is good qualitative agreement of the experimental data with the NPSPA\@. In 
particular, the temperatures of maximum local enhancement of the $C_V$ curves 
over a Fermi gas agree well between the experiment and the NPSPA\@.}
\label{fig2}
\end{figure}

As mentioned above, the SPA can avoid the sharp phase transition, which appears
in the simple BCS approximation. To explain this, we use the saddle-point 
approximation for the integral of Eq.\ (\ref{eq:2}), where we take into account
the measure $\Delta$ when the maximum of the integrand is determined. 
Neglecting the number projection for the sake of simplicity, the effective BCS 
approximation (EBCS) \cite{Rossignoli3} is obtained in this way. This leads to 
an effective value $\Delta_0$ which is determined by 
\begin{equation}
\Delta_0=G\langle0(\beta)|\hat{P}|0(\beta)\rangle+\frac{1}{2}GT/\Delta_0,
\label{eq:24}
\end{equation}
together with the condition for the particle number 
\begin{equation}
N=\sum_{k>0}\left[1-\left(u_k^2-v_k^2\right)\tanh(\beta\lambda_k/2)\right].
\label{eq:25}
\end{equation}
These equations are nonlinear and have to be solved by iteration in a similar 
way as the usual BCS equations. The solution of the above equations is a 
smooth, non-vanishing function of $T$\@. The second term in Eq.\ (\ref{eq:24}) 
prevents the solution from falling into the normal phase. Now, we introduce the
effective pairing gap as 
\begin{eqnarray}
\Delta_{\mathrm{E}}(T)=G\langle0(\beta)|\hat{P}|0(\beta)\rangle=\Delta_0-
GT/2\Delta_0.
\label{eq:26}
\end{eqnarray}
Here, one can see that if one neglects the second term in Eq.\ (\ref{eq:24}), 
the EBCS equation is reduced to the usual BCS equation. In this sense, the 
second term in Eq.\ (\ref{eq:24}) washes out the discontinuity of the heat 
capacity and avoids the vanishing pairing gap at the critical temperature in 
the BCS theory. Such a behavior is similar to the one seen in the BCS treatment
with number projection using the VAP scheme, where the sharp phase transition 
at the critical temperature is smoothed out. 

\begin{figure}[t!]
\includegraphics[totalheight=8.9cm]{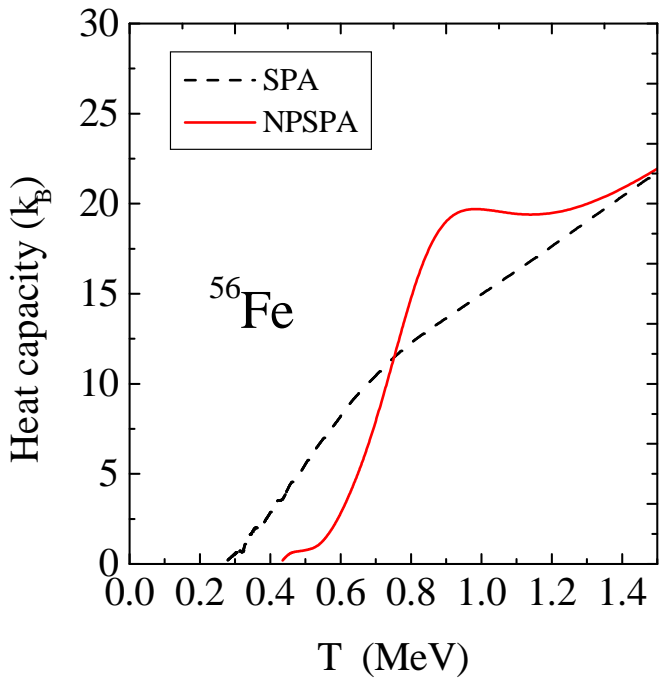}
\caption{(Color online) Heat capacity as a function of temperature for 
$^{56}$Fe. Solid and dashed curves as in Fig. \protect\ref{fig1}\@.}
\label{fig3}
\end{figure}

The expectation value $\langle0(\beta)|P_N|0(\beta)\rangle$ in Eq.\ 
(\ref{eq:16}) can now be evaluated using the $M$-point formula 
\begin{eqnarray}
\lefteqn{\langle0(\beta)|\hat{P}_N|0(\beta)\rangle=
\frac{1}{2\pi}\int_0^{2\pi}{\mathrm{d}}\,\varphi e^{i\varphi N}
\langle0(\beta)|e^{-i\varphi\hat{N}}|0(\beta)\rangle}\nonumber\\
&&\cong\frac{1}{M+1}\sum_{m=0}^Me^{i\varphi_mN}
\langle0(\beta)|e^{-i\varphi_m\hat{N}}|0(\beta)\rangle,\hspace*{1cm}
\label{eq:27}
\end{eqnarray}
where $\varphi_m=2\pi m/(M+1)$ and $M$ is the number of single particle states 
\cite{Hara}\@. In the case $M=1$, the number projection is reduced to the 
number-parity projection \cite{Rossignoli,Esashika}\@. 

\section{Results and Discussion}

In this work, we consider $^{172}$Yb, $^{94}$Mo, and $^{56}$Fe for numerical 
calculations. For these calculations, we use the single-particle energies 
$\varepsilon_k$ given by a deformed Woods-Saxon potential with spin-orbit 
interaction \cite{Cwoik}\@. The Woods-Saxon parameters are chosen such as to 
approximately reproduce the experimental single-particle energies extracted 
from the energy levels of $^{133}$Sn ($^{132}$Sn core plus one neutron) for 
$^{172}$Yb, where $V_0=51.0$~MeV, $a=0.67$~fm, $\kappa=0.67$, $\lambda=22.0$, 
and $r_0=1.27$~fm. For $^{94}$Mo and $^{56}$Fe, we adopt the same parameters 
as ones used in Ref.\ \cite{Kaneko2}\@. The doubly-degenerate single-particle 
levels with negative energies are taken outside of the $^{132}$Sn core for 
$^{172}$Yb, and outside of the $^{40}$Ca core for $^{94}$Mo and $^{56}$Fe. The
pairing force strengths are chosen such as to reproduce the experimental 
odd-even mass differences at zero temperature. 

\begin{figure}[t!]
\includegraphics[totalheight=9.1cm]{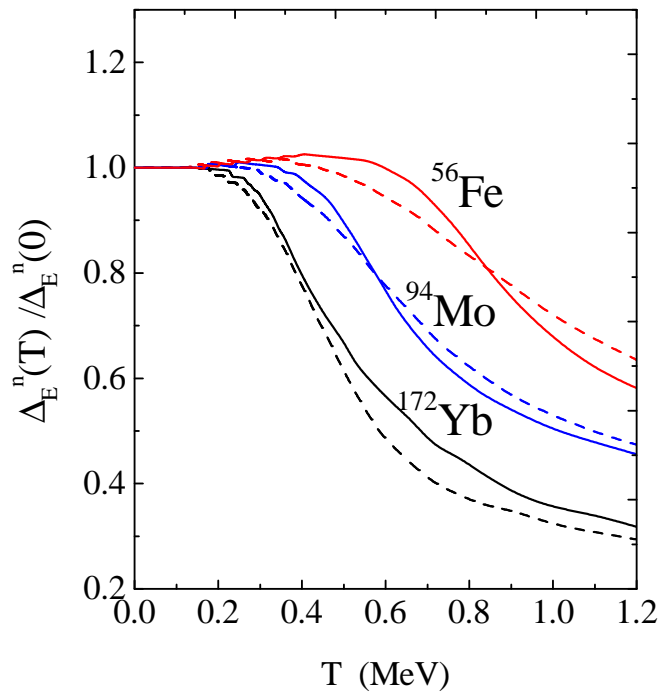}
\caption{(Color online) Effective pairing gap as a function of temperature for 
$^{172}$Yb, $^{94}$Mo, and $^{56}$Fe. Solid and dashed curves as in Fig.\ 
\protect\ref{fig1}\@.}
\label{fig4}
\end{figure}

Figure \ref{fig1} shows the heat capacities calculated in the SPA and 
number-projected SPA (NPSPA) for $^{172}$Yb. This heat capacity is obtained by 
$\partial E/\partial T$ where $E$ is the thermal energy given by 
$E=-\partial\ln Z_N/\partial\beta$\@. One can see that the heat capacity 
exhibits an $\cal{S}$ shape around $T=0.5$~MeV \cite{Schiller}\@. The number
projection decreases the SPA heat capacity for $T<0.35$~MeV and increases it in
the region of 0.35~MeV$<T<0.7$~MeV, and hence enhances the $\cal{S}$ shape. It 
is important to note that an $\cal{S}$-shaped heat capacity is already obtained
within the SPA, without any number projection. To emphasize this point, we also
show in Fig.\ \ref{fig1} the heat capacity obtained from an 
independent-particle model without pairing, i.e., for $G=0$\@. As it should for
a non-interacting Fermi gas, this heat capacity depends almost linearly on 
temperature. The difference between the heat-capacity curves of the 
independent-particle model and the full model treated within the SPA is large 
and qualitative. On the other hand, the difference between the SPA and NPSPA 
results is relatively smaller and quantitative in nature. When compared to 
experiment, the NPSPA can describe well the $\cal{S}$-shape of the data, while
the calculated $C_{v}$ curve deviates from the experimental one for 
temperatures below 0.5~MeV\@. This difference may be a drawback of the SPA 
approximation and of the simple model used with only a monopole pairing 
interaction and without any other interaction. As shown in Fig.\ \ref{fig2} for
$^{94}$Mo and in Fig.\ \ref{fig3} for $^{56}$Fe, however, the SPA does not 
produce an $\cal{S}$-shaped heat capacity by itself in those cases; the 
$\cal{S}$ shape appears only in the NPSPA\@. A recent analysis of poles in the 
complex temperature plane \cite{Schiller2} suggests that the pairing phase 
transition exists for mass $A>100$ but not for $A<100$\@. This interpretation 
seems to be consistent with the results obtained here. 

The $\cal{S}$ shape has also been discussed to be correlated with the breaking 
of nucleon Cooper pairs \cite{Liu,Kaneko1}\@. Therefore, we further investigate
the neutron pairing properties in the calculations\footnote{The neutron pairing
properties are very similar to the ones of proton pairing, therefore, proton 
pairing is not discussed separately in this work.}\@. In Fig.\ \ref{fig4}, we 
show the neutron effective pairing gap $\Delta_{\mathrm{E}}^n$ in the SPA and 
$\tilde{\Delta}_{\mathrm{E}}^n=G\langle0(\beta)|\hat{P}_N\hat{P}|0(\beta)
\rangle$ in the NPSPA, relative to their values at $T=0$\@. The suppression of 
$\Delta_{\mathrm{E}}^n$ is well correlated in temperature with the presence of 
the $\cal{S}$ shape of the heat capacity in Fig.\ \ref{fig1}\@, consistent with
the results of Refs.\ \cite{Liu,Kaneko1}\@. Thus, the $\cal{S}$ shape in Figs.\
\ref{fig1}, \ref{fig2}, and \ref{fig3} can be understood in terms of the 
suppression of the effective pairing gap and the effects of number projection. 
Previously, we have identified the inflection point of the effective 
pairing-gap curve as the temperature at which the pairing transition takes 
place \cite{Kaneko}\@. As seen from the respective curves for $^{172}$Yb in 
Fig.\ \ref{fig4}, this inflection point is close to the temperature 0.5~MeV at
which the heat-capacity curves peak\footnote{To obtain a precise estimate of 
the inflection point, we differentiate the effective pairing-gap curves with 
respect to temperature.}\@. For $^{94}$Mo and $^{56}$Fe, the respective 
inflection points of 0.65 MeV and 0.9~MeV are also close to the local maxima of
their $C_V$ curves. 

\begin{figure}[t!]
\includegraphics[totalheight=8.6cm]{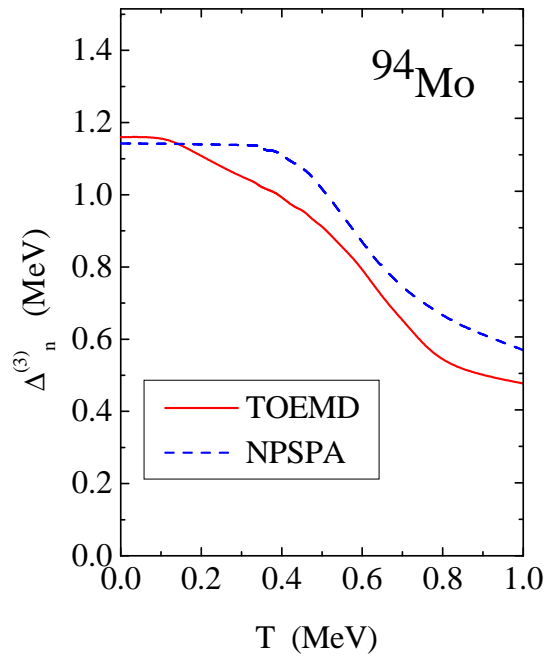}
\caption{(Color online) Comparison between the experimental TOEMD (solid line) 
extracted according to Eq.\ (\protect\ref{eq:28}) and the effective neutron 
pairing gap (dashed line) as a function of temperature for $^{94}$Mo.}
\label{fig5}
\end{figure}

As the experimental counterpart of the effective pairing gap, we have proposed 
in our previous work thermal odd-even mass differences (TOEMD) as a direct 
measure of the size of pairing correlations \cite{Kaneko}, and we have used 
them as indicators in our study of pairing phase transitions in $^{184}$W 
\cite{Kaneko1} and $^{94-97}$Mo \cite{Kaneko2}\@. We obtained in the one case a
drastic in the other case a gradual decrease of the TOEMD, and we found that 
this signal is well correlated with the $\cal{S}$ shape of the heat capacity. 
According to Ref.\ \cite{Kaneko}, the sudden decrease of the thermal odd-even 
mass differences is interpreted as a rapid breaking of nucleon Cooper pairs. 
Figure \ref{fig5} shows the comparison between the effective pairing gap and 
the TOEMD defined as 
\begin{eqnarray}
\lefteqn{\Delta_n^{(3)}(Z,N,T)=\frac{(-1)^N}{2}\left[B_t(Z,N+1,T)\right.}
\nonumber\\
&&\left.-2B_t(Z,N,T)+B_t(Z,N-1,T)\right],
\label{eq:28}
\end{eqnarray}
where the thermal energy $B_t$ is defined by $B_t(Z,N,T)=E(Z,N,T)+B(Z,N)$, 
$B(N,Z)$ is the binding energy at zero temperature, and $E(Z,N,T)$ is evaluated
from experimental level densities \cite{Kaneko1,Kaneko2}\@. The agreement 
between theory and experiment is satisfying. 

It is now interesting to discuss the significance of the $\cal{S}$ shape of the
heat capacity. For this reason, we would like to recall all the available 
facts. In the present work, the $\cal{S}$ shape is obtained for $A>100$ already
within the SPA, while for lighter nuclei, to reproduce the $\cal{S}$ shape, the
NPSPA (corresponding to a VAP scheme) is needed. Interestingly, in the work of 
Esashika \it et al.\ \rm \cite{Esashika}, an $\cal{S}$ shape could also be 
obtained (within a PAV scheme) when keeping the pairing gap $\Delta$ 
artificially constant. The latter two observations seem to indicate that the 
presence of the $\cal{S}$ shape is not necessarily related to the pairing phase
transition, and might be connected more to the particle-number projection. 
However, one should remember that in calculations without a distinct pairing 
force $G$, no $\cal{S}$-shaped heat capacity has been observed. Moreover, the 
temperature for a potential pairing phase transition derived from the 
$\cal{S}$-shaped heat capacity agrees very well with the temperature where 
pairing correlations are being suppressed (as indicated by the effective 
pairing gap \cite{Kaneko}), such that a simple coincidence of the two phenomena
can likely be ruled out. At present, we leave the question of the significance 
of the $\cal{S}$-shaped heat capacity somewhat open, while we would like to 
point out that the NPSPA is certainly an important tool to investigate this 
problem, since it enables us to obtain an $\cal{S}$-shaped heat capacity for 
lighter nuclei in the first place.

\section{Conclusion}

In conclusion, we have investigated effects of particle-number conservation in 
the SPA for $^{172}$Yb, $^{94}$Mo, and $^{56}$Fe. The particle-number 
projection affects the $\cal{S}$ shape of the heat capacity in all of these 
nuclei. The $\cal{S}$ shape in the heat capacity of $^{172}$Yb is produced by a
cooperation of the quantum effects in the SPA with the effects of 
particle-number projection. For $^{94}$Mo and $^{56}$Fe, however, the $\cal{S}$
shape of the heat capacity appears only in the calculation with particle-number
projection, but not in the SPA alone. This observation is consistent with 
nuclear size effects on the existence of a pairing transition. The 
$\cal{S}$-shaped heat capacity from NPSPA calculations correlates well in 
temperature with the reduction of the effective pairing gap. The effective 
neutron pairing gap in $^{94}$Mo is in good agreement with experimental thermal
odd-even mass differences. Our treatment of the particle-number projection can 
be applied together with the angular-momentum projection in order to study the 
spin distribution of nuclear levels in such a formalism as suggested in 
\cite{Tanabe}\@. Calculations are now in progress. 

\acknowledgments

Financial support from the National Science Foundation under Grant No.\ 
PHY-06-06007 is gratefully acknowledged. 



\end{document}